\documentclass[fleqn,twoside]{article}
\usepackage{espcrc2}

\usepackage{graphicx}
\usepackage[figuresright]{rotating}

\newcommand{\AmS}{{\protect\the\textfont2
  A\kern-.1667em\lower.5ex\hbox{M}\kern-.125emS}}

\title{Exploring the QCD phase diagram with compact stars}

\author{D. Blaschke \address[B]{Fakult\"at f\"ur Physik,
Universit\"at Bielefeld, D-33615 Bielefeld, Germany}
\address{Bogoliubov Laboratory for Theoretical Physics, JINR Dubna,
141980 Dubna, Russia}
        H. Grigorian\address[R]{Fachbereich Physik, 
Universit\"at Rostock, D-18051 Rostock, Germany}
        \thanks{On leave from: Department of Physics, 
Yerevan State University, 375025 Yerevan, Armenia}
        A. Khalatyan \address{Astrophysikalisches Institut Potsdam, 
          An der Sternwarte 16, D-14482 Potsdam, Germany}
        D. N. Voskresensky \address{Gesellschaft f\"ur 
          Schwerionenforschung mbH, Planckstr. 1, D-64291 Darmstadt, Germany}
 \thanks{Permanent address: Moscow Institute for Physics and Engineering, 
 115409 Moscow, Russia}}
\begin{document}

\begin{abstract}
We investigate a nonlocal chiral quark model with separable 
4-fermion interaction for the case of $U(3)$ flavor symmetry
and show that strange quark matter is unlikely to occur in 
a large enough volume of a compact star to entail remarkable
observational consequences. The phase diagram in the two-flavor 
sector of such model has a critical end point of the line of first order 
chiral/deconfinement phase transitions on which a triple point marks 
the junction with the critical line for second order phase transitions to 
two-flavor color superconductivity (2SC) below $T\sim 80$ MeV.
Stable hybrid star configurations with large quark matter core 
in a color superconducting phase can exist. 
A consistent cooling phenomenology requires that all quark species be 
gapped, the minimal pairing gap of the order of $10 - 100$ keV. 
\vspace{1pc}
\end{abstract}

\maketitle

\section{Introduction}
The investigation of the QCD phase diagram 
has become a research topic of highest priority. 
Relativistic heavy ion collisions in the RHIC era have provided results
which confirm previous information about the critical temperature 
$T_c\sim 170$ MeV of the hadronization transition in the 
approximately baryon-free regime
\cite{Braun-Munzinger:2003zd}, which 
parallels the situation a few microseconds after the big bang.    
This value is in  agreement with the deconfinement transition 
temperature calculated in 2+1 flavor Lattice QCD
\cite{Karsch:2003jg}.
These simulations are now extended into the region of (small) finite 
chemical potentials where one of the characteristic features is a critical 
endpoint of first order phase transitions (tricritical point).    
Future heavy-ion collision experiments planned at GSI Darmstadt (FAIR)
will explore the phase diagram in the finite density domain 
and hope to find experimental signatures of the tricritical point.
In dense quark matter at temperatures below $\sim 50$ MeV, due to
attractive interaction channels, the Cooper pairing instability is 
expected to occur which should lead to a variety of possible quark
pair condensates corresponding to color superconductivity (CSC) phases,
see \cite{Alford:2001dt} for a review.
Since it is difficult to provide low enough temperatures for CSC phases
in heavy-ion collisions, only precursor phenomena 
\cite{Kitazawa:2001ft,Voskresensky:2004jp} are expected under these conditions.
CSC phases may occur in neutron star interiors \cite{nsi} and 
could manifest themselves, e.g., in the cooling behavior 
\cite{Blaschke:1999qx,
Blaschke:2004vr}. 

However, the domain of the QCD phase diagram where neutron star conditions
are met is not yet accessible to Lattice QCD studies and theoretical 
approaches have to rely on nonperturbative QCD modeling. 
The class of models closest to QCD are Dyson-Schwinger equation 
(DSE) approaches which have been extended recently to finite temperatures and 
densities \cite{Bender:1996bm
}.
Within simple, infrared-dominant DSE models early studies of quark stars
\cite{Blaschke:1998hy} and  diquark condensation \cite{Bloch:1999vk} have 
been performed.

The present contribution will be based on results which have been obtained
within a nonlocal chiral quark model (NCQM) using a covariant, separable 
representation of the gluon propagator \cite{Burden:1996nh} with formfactors
suitable for an extension to finite temperatures and chemical potentials
\cite{Blaschke:1998gk,Blaschke:1999ab,Blaschke:2000gd}.
Alternative approaches to separable chiral quark models are based on the 
instanton approach \cite{separable,sepgauss,GomezDumm:2001fz}.
As a limiting case, the well-known Nambu--Jona-Lasinio (NJL) model is 
obtained for a cutoff formfactor. 
For recent applications of the NJL model to QCD at
high density in compact stars, see \cite{Buballa:2003qv} and references 
therein.
A different class of QCD-models is of the bag-model type. These models do
not describe the chiral phase transition in a self-consistent way and will
not be considered here.
We will consider within a NCQM the following questions: 
(i) Is strange quark matter relevant for structure and evolution of 
compact stars? 
(ii) Are stable hybrid stars with quark matter interior possible?
(iii) What can we learn about possible CSC phases from neutron star cooling?

\section{Nonlocal chiral quark model, $N_f=3$}

In order to answer the question whether strange quark matter phases 
should be expected in the neutron star interiors, we consider here 
a three-flavor generalization of a NCQM with the action 
%
\begin{eqnarray}
  \label{eq:4}
 && {\cal S}[q, \bar q] =
\int_k\big\{\bar q(k) (\! \not k - \hat m)q(k)
              \nonumber \\
           &&+ D_0 \int_{k^\prime}\sum_{\alpha=0}^{8} 
                 [j_s^\alpha(k)j_s^\alpha(k^\prime) 
+ j_p^\alpha(k)j_p^\alpha(k^\prime)]\big\}, 
\end{eqnarray}
where  $\int_k=\int \frac{d^4\!k}{(2\pi)^4} $. 
We restrict us here to the {scalar} current 
$j_s^\alpha(k)=\bar q(k)\lambda_\alpha f(k)q(k)$
and the pseudoscalar current 
$j_p^\alpha(k)=\bar q(k)i\gamma_5\lambda_\alpha f(k)q(k)$
in Dirac space with $q(k)$ and $\bar q(k)$ being quark 4-spinors and the 
formfactor $f(k)$ accounts for the nonlocality of the interaction.
For the applications to compact stars, the quark matter
equation of state is needed and can be obtained from the 
partition function 
%
\begin{equation}
  \label{eq:5}
  {\cal Z}[T,\hat\mu] = \int {\cal D}\bar q {\cal D}q \exp\left(
         {\cal S}[q, \bar q]  - \int_k 
 \hat\mu\gamma^0\bar qq \right),
\end{equation}
where the constraint of baryon number conservation is realized by the diagonal
matrix of chemical potentials $\hat\mu$ (Lagrange multipliers). 
Further details concerning the notation and the
parametrization of this model can be found in Ref. \cite{GBKG}.

In order to perform the functional integrations over the quark fields 
$\bar q$ and $q$ we use the formalism of bosonisation  
which is based on the Hubbard-Stratonovich transformation 
of the four-fermion interaction. The resulting transformed partition function 
in terms of bosonic variables will be considered in the mean-field 
approximation
\begin{eqnarray}  \label{eq:6}
{\cal Z}[T,\hat\mu]= 
{\rm \large e}^{\large -2N_c \sum_f\bigg[\frac{\phi_f^2}{8D_0}
 + \int_k \ln(M_f^2-\tilde k_f^2) \bigg]},
\end{eqnarray}
with the 4-vector $\tilde k_f = (k_0-\mu_f,{\vec k})$
and the effective quark masses 
$M_f=M_f(\tilde k)=m_f + \phi_f f(\tilde k)$
containing the flavor dependent mass gaps $\phi_f$. 
In the mean field approximation, the grand canonical thermodynamical potential 
is
\begin{eqnarray}
  \label{eq:9}
\Omega(T,\{\mu\}) = \frac{T}{V}\ln\left\{{\cal Z}[T,\{\mu_f\}]/{\cal
    Z}[0,\{0\}]\right\}~,
\end{eqnarray}
where fluctuations are neglected and the divergent vacuum contribution 
has been subtracted.
The quark mass gaps $\phi_f$ are determined by 
solving the gap equations which correspond to the minimization conditions
${\partial\Omega}/{\partial\phi_f}=0$ and read
\begin{equation}
  \label{eq:14}
  \phi_f = 8D_0 N_c \int\frac{d^4\!k}{(2\pi)^4}\frac{2M_ff(\tilde k)}
{M_f^2-\tilde k_f^2}~.
\end{equation}
As can be seen from (\ref{eq:14}), for the chiral $U(3)$ quark model 
the three gap equations for $\phi_u,~\phi_d,~\phi_s$ are decoupled
and can be solved separately. 

In what follows we consider the case $T=0$ only. 
All thermodynamical quantities can now be derived from Eq.\ (\ref{eq:9}),
for details see \cite{GBKG}. 
For instance,
pressure, density and the chiral condensate are given by
\begin{eqnarray}
  \label{eq:13}
  p = -\Omega\;,\;n_f = -\frac{\partial\Omega}{\partial\mu_f}\;,\;
<\!\bar q q\!> = -\frac{\partial\Omega}{\partial m_q}~,
\end{eqnarray}
the energy density follows from $\varepsilon = -p+\sum_f\mu_f n_f$.
%
For most of the numerical investigations within the NCQM described above  
we will employ a simple Gaussian formfactor 
$f(k)=\exp(k^2/\Lambda^2)~,$
which has been used previously for studies of meson 
and baryon  properties in the vacuum \cite{sepgauss} as well as quark 
deconfinement and mesons at finite temperature 
\cite{Blaschke:1998gk,Blaschke:2000gd}. 
A systematic extension to other choices of formfactors can be found in
\cite{GomezDumm:2001fz}.
The Gaussian model has four free parameters to be defined: the coupling  
constant $D_0$, the interaction range $\Lambda$, and the current quark
masses $m_u=m_d$, $m_s$.
These are fixed by the pion mass $m_\pi=140$ MeV, kaon mass $m_K=494$ MeV,
pion decay constant $f_\pi=93$ MeV and the chiral condensate
$-<\!\bar q q\!>= (240~{\rm MeV})^4$. 
%
%
\begin{figure}[htb]
   \vspace*{-0.5cm}
    \includegraphics[width=0.9\linewidth]{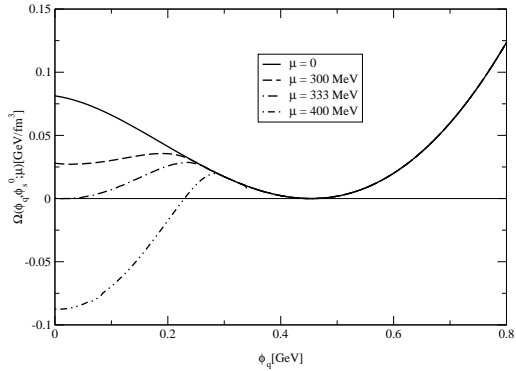} 
   \vspace*{-0.8cm}
    \caption{Dependence of the thermodynamical potential on the light flavor 
        gap $\phi_q=\phi_u=\phi_d$
        (order parameter) for different values of the chemical potential,
        $\phi_s=682$ MeV.
        }
    \label{fig1}
   \vspace*{-0.5cm}
\end{figure}
Fig.\ \ref{fig1} shows the behavior of the thermodynamical potential 
as a function of the light quark mass gap $\phi_q=\phi_u=\phi_d$ for different 
values of the chemical potential $\mu$. 
For $\mu<\mu_c=330$ MeV the argument and the value of the global minimum is 
independent of $\mu$ which corresponds to a vanishing quark density 
(confinement). 
At $\mu=\mu_c$ a phase transition from the 
massive, confining phase to a deconfining phase with negligibly small mass 
gap occurs. 
\begin{figure}[htb]
    \includegraphics[width=0.9\linewidth]{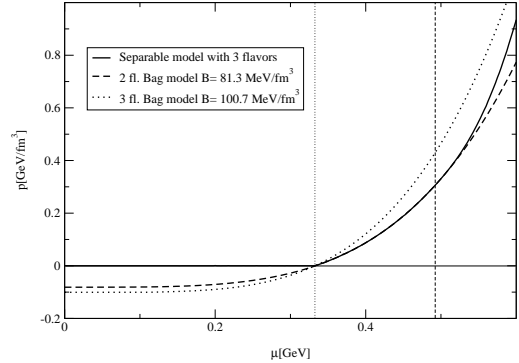} 
    \vspace*{-0.8cm}
    \caption{Pressure of the NCQM as a function of the chemical 
potential for the separable model (solid line) compared to a three-flavor 
(dotted line) and a two-flavor (dashed line) bag model. All models have the
same critical chemical potential $\mu_c=330$ MeV for (light) quark 
deconfinement.}
    \label{fig3}
   \vspace*{-0.5cm}
\end{figure}
In Fig.\ \ref{fig3} we show that the quark 
pressure vanishes for $\mu<\mu_c$ (confinement) and can be well described
by a two-flavor bag model with a bag constant $B=81.3$ MeV/fm$^3$ in the 
region of chemical potentials $330$ MeV $\le \mu \le 492$ MeV where the third 
flavor is still confined.
\section{Color superconductivity, $N_f=2$}
The quark-quark interaction in the color anti-triplet scalar diquark 
channel is attractive driving the pairing with a large zero-temperature
pairing gap $\Delta\sim 100$~MeV for the quark chemical
potential $\mu_q \sim 300-500$~MeV. Therefore, we consider now a NCQM
described by the effective action
\begin{eqnarray}
S_{2sc}[\bar q, q]&=&S[\bar q, q]+\frac{3D_0}{4}
\int_{k,k'}j_d^\dagger(k)j_d(k'),
\end{eqnarray}
where $j_d(k)=q^{T}(k) C i\gamma_5\tau_2\lambda_2 f(k) q(k)$
is the scalar diquark current with the  charge conjugation matrix
$C=i\gamma_0\gamma_2$.
After bosonization, the mean field approximation
introduces a new order parameter: the diquark gap
$\Delta$, which can be seen as the gain in energy due to diquark
condensation. The mass gaps $\phi_u$, $\phi_d$  indicate
dynamical chiral symmetry breaking. The grand canonical thermodynamic
potential per volume can be obtained as
\begin{eqnarray}\label{thpot}
&&\Omega_q(\phi_u,\phi_d,\Delta;\mu_u,\mu_d,T)=
\frac{\phi_u^2+\phi_d^2}{8~D_0}+\frac{|\Delta|^2}{3~D_0}\nonumber\\
&&-\frac{T}{2}\sum_n\!\int\!
\frac{d^3k}{(2\pi)^3}{\rm Tr}\ln\left[\frac{1}{T}
G^{-1}(i\omega_n,\vec{k})\right],
\end{eqnarray}
where $\omega_n=(2n+1)\pi T$ are the Matsubara frequencies for
fermions and the inverse Nambu-Gorkov quark propagator is 
\begin{equation}
G^{-1}= \left(
\begin{array}{cc}
\not\!k - \hat M -\hat{\mu}\gamma_0~~~&
\Delta \gamma_5\tau_2\lambda_2 f(k)\\
-\Delta^\ast \gamma_5\tau_2\lambda_2 f(k)& \not\!k - \hat
M+\hat{\mu}\gamma_0
\end{array}
\right)~.
\end{equation}
\begin{figure}[ht]
  \vspace{-1cm}
    \hspace{-0.05\linewidth}
    \includegraphics[width=0.9\linewidth,angle=-90]{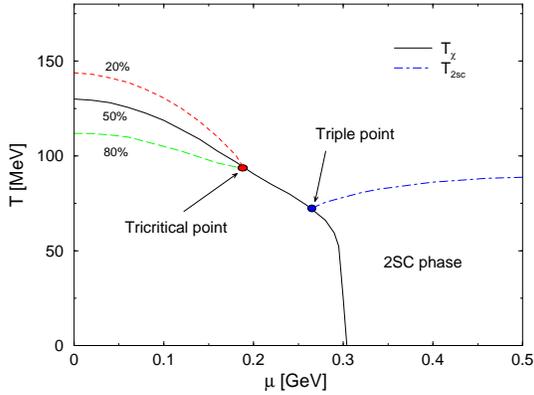}
  \vspace{-0.8cm}
  \caption{Phase diagram for $N_f=2$ quark matter in the NCQM.
 \label{fig:phd}}
 \vspace*{-0.5cm}
\end{figure}
The resulting phase diagram is shown in Fig. \ref{fig:phd}.
It includes a 2-flavor color
superconductivity (2SC) phase for which quarks of one color, say blue,
remain unpaired.  The color-flavor locking 
(CFL) phase \cite{arw98} requires approximate SU(3) flavor symmetry 
and can be excluded from our discussion since strange quarks remain
confined up to the highest densities occuring in a compact star
configuration \cite{GBKG}.

\section{Structure and cooling of hybrid stars}

In describing the hadronic shell of a hybrid star we  exploit a parametrized 
form \cite{HJ99} of
the Argonne $V18+\delta v+UIX^*$ model of the EoS given in \cite{APR98}, which
is based on the most recent models for the nucleon-nucleon interaction with the
inclusion of a parameterized three-body force and relativistic boost 
corrections. All details concerning the EoS and the hadronic cooling 
processes are given in \cite{Blaschke:2004vq}.
As we have seen in the previous section, it may be sufficient to discuss only 
two-flavor quark matter for applications to compact stars.
We will focus on the model of the quark EoS developed in
\cite{BFGO} as an instantaneous approximation to the covariant
NCQM (INCQM). 
\begin{figure}[ht]
    \includegraphics[width=0.9\linewidth,angle=-90]{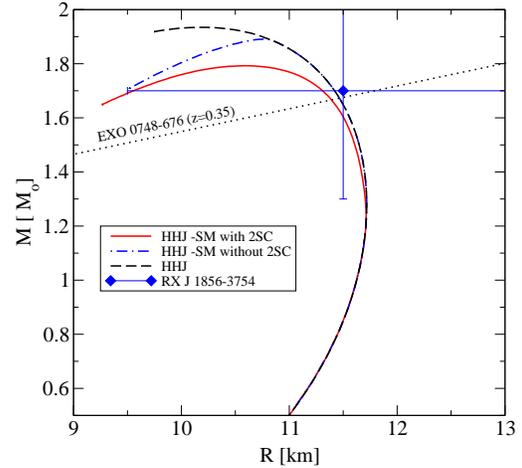}
\vspace{-0.8cm}
\caption{Mass - radius 
relations for compact star configurations with different EoS:
purely hadronic star with HHJ EoS (dashed),
stable hybrid stars for HHJ - INCQM$_{\rm Gauss}$ EoS with 2SC 
(solid) and without 2SC phase (dash-dotted).
For comparison, observational constraints on the compactness are given 
from the "small" compact star RXJ 1856 and from the high surface redshift
object EXO 0748 which can both be obeyed by our hybrid star EoS. 
\label{fig:stab}}
\vspace{-0.5cm}
\end{figure}

In Refs. \cite{VYT02} it has been demonstrated on the example of
the hadron-quark mixed phase that finite size effects might play 
a crucial role by substantially narrowing the
region of the mixed phase or even preventing its appearance.
Therefore we omit the possibility of the hadron-quark mixed phase in our
model where the quark phase arises by the Maxwell construction.
For the case of the instantaneous Gaussian formfactor with 2SC phase
the quark core appears for $M>1.21~M_\odot$. Without 2SC phase or for 
the Lorentzian or NJL formfactor no stable hybrid stars are obtained,
see Fig. \ref{fig:stab}. 

\begin{figure}[htb]
\vspace{-0.5cm}
 \includegraphics[width=0.9\linewidth,angle=-90]{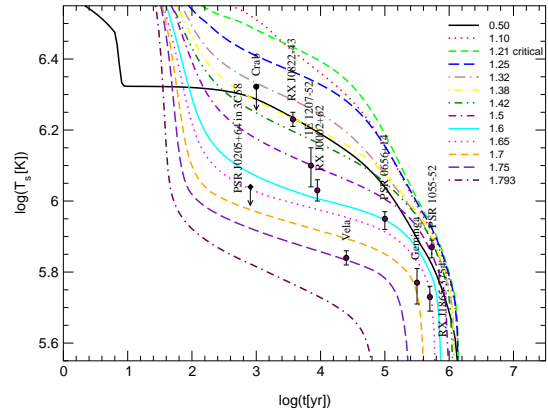}
\vspace{-0.8cm}
\caption{Cooling of hybrid star configurations of Fig. \ref{fig:stab} 
with color superconducting quark matter core in 2SC+X phase.
Different lines correspond to hybrid star masses in units of the solar
mass.}
\label{fig:cool-csl-x}
\vspace{-0.5cm}
\end{figure}

In order to describe the cooling of stable hybrid star configurations,
we use the approach to the hadronic cooling developed recently in Ref. 
\cite{Blaschke:2004vq} and add the contribution due to the quark core,
see \cite{Blaschke:2004vr}.
Once in the quark matter core of a hybrid star unpaired quarks are
present the dominant quark direct Urca (QDU) process is operative and 
cools the star too fast in disagreement with the observational data 
\cite{Blaschke:2004vr}. 
If one allows for a residual weak pairing channel as, e.g., 
the CSL one with typical gaps of $\Delta \sim 10$~keV - $10$~MeV, 
see \cite{Schmitt:2004hg}, the hybrid star configuration will not be
affected but the QDU cooling process will be efficiently 
suppressed as is required for a successful description of compact star 
cooling.
Since we don't know yet the exact pairing pattern for which also the
residual (in 2SC unpaired) quarks are paired in the 2-flavor quark matter, 
we will call this hypothetical phase ``2SC+X''.
In such a way all the quarks may get paired, either strongly in the 
2SC channel or weakly in the X channel.
If now the X-gap is of the order of $1$ MeV, then the QDU process will
be effectively switched off and the cooling becomes too
slow, again in disagreement with the data \cite{Blaschke:2004vr}.
This seems to be a weak point and we would like to explore whether a
density-dependent X-gap could allow a description of the cooling data 
within a larger interval of compact star masses. 
For such a density-dependent X-gap function we employ the ansatz
\begin{equation}
\label{gap}
\Delta_X(\mu)=\Delta_c~\exp[-\alpha(\mu-\mu_c)/\mu_c]~.
\end{equation}
In Fig. \ref{fig:cool-csl-x} we show the resulting cooling curves for 
the parameters $\Delta_c=1.0$ MeV and $\alpha=10$. 
%
We observe that the mass interval for compact stars which obey the 
cooling data constraint ranges now from $M=1.32~M_\odot$ for slow
coolers up to $M=1.75~M_\odot$ for fast coolers such as Vela.

\section{Conclusion}
We have presented the QCD phase diagram obtained within the NCQM approach.
We find that in the framework of our model for neutron star
applications it is sufficient to consider
only two quark flavors since the critical densities of strange quark 
deconfinement are not accessible in typical compact star interiors.
We investigate the phase transition between superconducting quark matter 
and hadronic matter under neutron star constraints and find 
that stable hybrid stars are possible
with large quark matter cores in the 2SC phase.
The solution of the cooling evolution
shows that the presence of ungapped quark species entails too fast
cooling due to the QDU process in disagreement with modern neutron star
cooling data.
A successful description of hybrid star cooling requires a pairing pattern 
in the quark matter core (if it exists) where all quarks are gapped and
a weak pairing channel is present with a gap which does not exceed 
about $10-50$ keV with a decreasing density dependence. 
A precise microscopic model for this conjectured "2SC+X" phase is still
lacking. Possible candidate for a weak pairing channel is the color-
spin-locking (CSL) phase.  
We have demonstrated on the example of hybrid star cooling how astrophysical 
observations could provide constraints for microscopic approaches to the 
QCD phase diagram in the domain of cold dense matter not accessible to 
heavy-ion collision experiments and lattice QCD simulations.
\section*{Acknowledgments}
We like to thank our colleagues D.N. Aguilera, M.\ Buballa, 
S. Fredriksson, C. Gocke, Yu.\ Kalinovsky, A. \"Ozta\c{s}, F. Sandin, 
N.\ Scoccola and P.C.\ Tandy for their contributions to the
formulation of separable quark models. 
D.B. thanks for partial support of the Department of 
Energy during the program INT-04-1 on {\it QCD and Dense Matter: 
From Lattices to Stars} at the University of Washington,
D.N.V. was supported by DFG grant 436 RUS 113/558/0-2.
We acknowledge the support of DAAD for scientist exchange between the 
Universities of Rostock and Yerevan.  
This research has been supported in part by the Virtual Institute 
{\it Dense hadronic matter and QCD phase transition} of the
Helmholtz Association under grant number VH-VI-041.

\end{document}